\documentclass[conference]{IEEEtran}
\IEEEoverridecommandlockouts

\usepackage{cite}
\usepackage{amsmath,amssymb,amsfonts}
\usepackage{graphicx}
\usepackage{textcomp}
\usepackage{xcolor}
\usepackage{colortbl}
\usepackage{todonotes}
\usepackage{enumerate}
\usepackage{url}
\usepackage{tcolorbox}
\usepackage{tikz}
\usepackage[framemethod=TikZ]{mdframed}
\definecolor{shadecolor}{RGB}{163, 210, 186}

\def\BibTeX{{\rm B\kern-.05em{\sc i\kern-.025em b}\kern-.08em
    T\kern-.1667em\lower.7ex\hbox{E}\kern-.125emX}}
    
\makeatletter
\newcommand{\linebreakand}{%
  \end{@IEEEauthorhalign}
  \hfill\mbox{}\par
  \mbox{}\hfill\begin{@IEEEauthorhalign}
}
\makeatother   
\begin{document}

\title{Enhancing Women's Experiences in Software Engineering}

\author{\IEEEauthorblockN{Júlia Rocha Fortunato}
\IEEEauthorblockA{\textit{Faculty UnB Gama (FGA)} \\ 
\textit{University of Brasilia (UnB)} \\
Brasilia, Brazil \\
juliarochafort@gmail.com}

\and

\IEEEauthorblockN{Luana Ribeiro Soares}
\IEEEauthorblockA{\textit{Faculty UnB Gama (FGA)} \\ 
\textit{University of Brasilia (UnB)} \\
Brasilia, Brazil \\
luana.soares0901@gmail.com}

\and

\IEEEauthorblockN{Gabriela Silva Alves}
\IEEEauthorblockA{\textit{Faculty UnB Gama (FGA)} \\ 
\textit{University of Brasilia (UnB)} \\
Brasilia, Brazil \\
gabrielaalves.gsa@gmail.com}

 \linebreakand

\IEEEauthorblockN{Edna Dias Canedo}
\IEEEauthorblockA{\textit{Department of Computer Science} \\ 
\textit{University of Brasilia (UnB)} \\
Brasilia, Brazil \\
ednacanedo@unb.br}

\and

\IEEEauthorblockN{Fabiana Freitas Mendes}
\IEEEauthorblockA{\textit{Department of Computer Science} \\
\textit{Aalto University}\\
Espoo, Finland \\
fabiana.mendes@aalto.fi}
}

\maketitle

\begin{abstract}
    \textbf{Context:} 
    Women face many challenges in their lives, which affect their daily experiences and influence major life decisions, starting before they enroll in bachelor's programs, setting a difficult path for those aspiring to enter the software development industry.
    \textbf{Goal:}  
    To explore the challenges that women face across three different life stages—beginning as high school students, continuing as university undergraduates, and extending into their professional lives—as well as potential solutions to address these challenges.
    \textbf{Research Method:}
    We conducted a literature review followed by workshops to understand the perspectives of high school women, undergraduates, and practitioners regarding the same set of challenges and solutions identified in the literature.
    \textbf{Results:} 
    Regardless of the life stage, women feel discouraged in a toxic environment often characterized by a lack of inclusion, harassment, and the exhausting need to prove themselves. We also discovered that some challenges are specific to certain life stages; for example, issues related to maternity were mentioned only by practitioners.
    \textbf{Conclusions:} 
    Gender-related challenges arise before women enter the software development field when the proportion of men and women is still similar. While the need to prove themselves is mentioned at all three stages, high school women's challenges are more often directed toward convincing their parents that they are mature enough to handle their responsibilities. As they progress, the emphasis shifts to proving their competence in managing responsibilities for which they have received training. Increasing the inclusion of women in the field should, therefore, start earlier, and profound societal changes may be necessary to boost women's participation.
\end{abstract}

\begin{IEEEkeywords}
Gender diversity, Gender barriers and challenges, life stages, software engineering  
\end{IEEEkeywords}

\section{Introduction} \label{sec:intro}

Women encounter many challenges at every stage of their lives that collectively impact their opportunities in many fields, including software engineering \cite{canedo_breaking_2021}. From childhood through adulthood, these challenges are often rooted in societal norms, stereotypes, and systemic biases that create barriers to entry and success in technology-related careers.

In childhood, girls are frequently discouraged from pursuing interests in science, technology, engineering, and mathematics (STEM) due to gender stereotypes. The lack of women role models in these fields and the nature of toys and activities marketed to girls further diminish their exposure to and interest in technology. As they transition into adolescence, social pressures intensify, and girls may find themselves away from STEM  subjects due to unequal educational support and limited encouragement to engage in tech-related activities like coding or robotics \cite{de2023autopercepccao,mcguire2022gender,rogers2021school,martin2021if, king2021gender,gonzalez2020girls,reinking2018gender,DBLP:conf/iticse/WangHRI15,DBLP:journals/jiteiip/SullivanB16}.

By young adulthood, the cumulative effect of these earlier experiences often results in fewer women choosing technology-oriented university programs. For those who do, the university environment can be isolating and discriminatory, further deterring their progress. As they enter the professional world in adulthood, women face additional challenges such as balancing work and family responsibilities, navigating gender discrimination, and coping with a lack of mentorship and career advancement opportunities \cite{tokbaeva2023career,DBLP:conf/sbes/CanedoCSM22,DBLP:journals/tosem/TrinkenreichWSG22,dias_canedo_barriers_2019,canedo_breaking_2021,o2020so}.

These challenges, while manifesting differently at each life stage, represent a continuous thread that influences women's career trajectories in technology \cite{kohl_challenges_2021,trinkenreich_empirical_2022,DBLP:conf/icse/BreukelenBBS23,DBLP:conf/ge-ws/Canedo0LSM24}. To significantly increase the representation of women in software engineering, it is essential to address these challenges from the very beginning of their lives. By challenging stereotypes, promoting STEM engagement from an early age, and creating supportive and inclusive environments, we can work towards breaking down these barriers \cite{DBLP:conf/amcis/AufschlagerKWKW23,trinkenreich_empirical_2022,DBLP:conf/csci/KramarczukAPN21,dias_canedo_barriers_2019}.

Despite the strides made in gender equality throughout the 20th century, the software engineering field remains predominantly men, with women still facing substantial discrimination and underrepresentation \cite{DBLP:conf/icse/BomanAN24,DBLP:conf/icse/OliveiraBSBBF24}. The persistence of gender bias, lack of recognition, and limited opportunities for advancement highlight the ongoing struggle for equality in this sector\cite{soares_investigating_2023}. Moreover, the gender gap in software engineering not only limits diversity but also deprives the industry of valuable perspectives that could drive innovation and growth \cite{DBLP:conf/icse-chase/FengGGS23, DBLP:journals/tosem/TrinkenreichWSG22,DBLP:journals/computer/TrinkenreichGS22, canedo_breaking_2021}.

This research aims to \textbf{explore the challenges that women face across three different life stages—beginning as high school students, continuing as university undergraduates, and extending into their professional lives—as well as potential solutions to address these challenges}. By bringing academic knowledge to these groups and fostering discussions around these issues, we can raise awareness and create more targeted interventions to support women in pursuing careers in software engineering. Understanding the evolving nature of these challenges is critical to developing effective strategies that can attract and retain more women in this field, ultimately contributing to a more diverse and equitable industry.

\section{Related Work} 
\label{sec:related}

Research has identified significant challenges affecting women's participation in IT and software engineering, beginning in high school and extending into their professional careers. Early in their education, both girls and boys rate their abilities in STEM subjects similarly, but as they grow older, girls increasingly doubt their skills, particularly in IT \cite{tomte2011gender, tomperi2022investigation, kaarakainen2019information}. This decline in self-assessment impacts career decisions, as demonstrated by Hinckle et al. \cite{hinckle2020relationship}. Personal experiences also play a key role, as highlighted by Silvast \cite{silvast2015oral}.

Vainionp{\"a}{\"a} et al.  \cite{vainionpaa2020girls} conducted a study combining literature reviews and interviews, identifying intentional and unintentional factors influencing girls' decisions about IT careers. These include the pressure of university entrance exams, stereotypes, and the influence of family members. The authors categorized women’s attitudes toward IT careers into three groups: resistant, indifferent, and explorer \cite{vainionpaa2020not}. Resistant individuals often lacked confidence, while explorers were more open to pursuing IT. Similarly, Tahsin et al. \cite{DBLP:conf/icse/TahsinAAS22} explored the underrepresentation of women in IT in Bangladesh, identifying barriers such as societal stigma, parental influence, lack of career counseling, and security concerns. Solutions included mentoring programs and initiatives to address stereotypes, educate parents, and highlight women’s achievements in ICT.

University experiences often exacerbate challenges. Fietta et al. \cite{DBLP:conf/goodit/FiettaNMG23} found that women in Computer Science felt undervalued and faced gender bias from peers and professors, leading to emotional discomfort. Additionally, Perdriau et al. \cite{DBLP:conf/sigcse/PerdriauOGLL24} reported that the ``diversity-hire” narrative perpetuated by peers and family members increased self-doubt among women.

In professional environments, workplace culture often reflects similar biases. Oliveira et al. \cite{DBLP:conf/icse/OliveiraBSBBF24} found that gender bias, harassment, impostor syndrome, and lack of inclusion persist from academia into the workforce. These issues lead to low retention rates and limited career advancement for women. Implicit biases further marginalize women, as Wang \cite{wang_implicit_2019} demonstrated, showing how they disadvantage women software engineers.
Canedo et al. \cite{canedo_breaking_2021} also discussed the impact of stereotypes, impostor syndrome, and workplace toxicity on women’s confidence and career progression.

This research builds on these studies to investigate the challenges faced by women in IT and software engineering from high school through their professional careers, proposing solutions to address these barriers.

 \section{Research Method} \label{sec:meth}

This work was conducted as part of the \textbf{Protagonists in Software Engineering (PES)} project, led by three women pursuing their bachelor's degrees in Software Engineering. The research goals are presented following.

\begin{enumerate} [RG 1. ]
    \item  Understand and categorize the challenges women face in the Software Engineering job market.\\
    \textit{Motivation: Although there has been research on the challenges women face in this field, there is a lack of comprehensive reviews that document and aggregate them in comprehensive categories.}
        
    \item Disseminate knowledge about these challenges and their potential solutions.\\
    \textit{Motivation: Awareness is the first step in addressing any issue. Without awareness, it is impossible to tackle the problem. This goal involves discussing the challenges and solutions identified in the literature with women at different stages of their careers.}
    
    \item Report the findings and experiences gained through the PES project.\\
    \textit{Motivation: share and spread knowledge on the subject to improve the situation for women in Software Engineering.}
\end{enumerate}

\begin{figure}[htbp]
    \centering \includegraphics[scale=0.15]{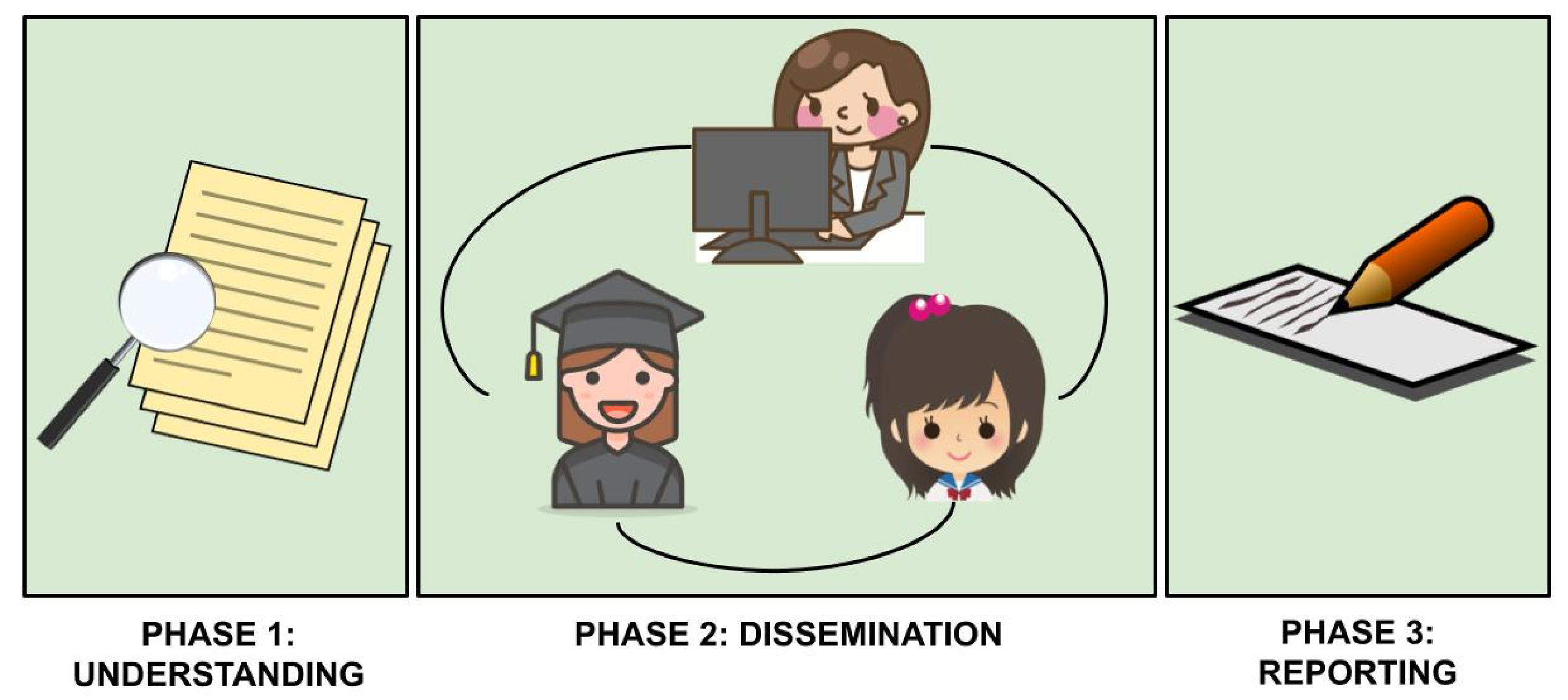}
    \caption{Research Phases}
    \label{fig:research-phases}
\end{figure}

The research was structured into three phases, each corresponding to one of the research goals: Understanding, Dissemination, and Reporting, as illustrated in Figure \ref{fig:research-phases}.

During the \textbf{Understanding Phase}, we conducted a literature review to synthesize the challenges and solutions related to gender equality in Software Engineering. To guide our literature review, we defined two research questions: \textbf{RQ 1.} What challenges do women face when working in software engineering roles? and \textbf{RQ 2.} Are there defined actions to overcome the reported challenges?

Based on the literature review findings, during the \textbf{Dissemination Phase}, we organized workshops to discuss the literature review results with three different groups of women: (1) students from a public high school, (2) the university community (professors and students), and (3) software engineering practitioners. The workshops were conducted separately and sequentially. Each workshop followed the same structure: I. Introduction to the project; II. A word cloud activity using the Mentimeter website, where participants contributed words representing the challenges faced by women; III. A discussion of the word cloud results; IV. Presentation of the challenges identified in the literature; V. Solutions workshop where participants selected one of the discussed challenges and propose a solution; VI. Presentation of solutions found in the literature, followed by a discussion comparing the literature-based solutions with those proposed by the participants.

Finally, in the \textbf{Reporting Phase}, we produced documents and presented the results to communities interested in this subject. The following sections provide a detailed overview of each phase.

\section{Challenges and Solutions}\label{sec:chall}

Our study investigated women's challenges at various stages of their journey into the software development field, from high school to professional practice. We identified recurring challenges that persist throughout their lives and some specific to particular stages. In order to identify these patterns, we compared the challenges reported in the literature with those collected during our workshops, as illustrated in Table \ref{tab:comp}.

The first column of Table \ref{tab:comp} lists the categories of challenges identified in the literature, followed by the studies from where we collected them. The subsequent columns compare these with the findings from the workshops. Although there are slight differences in interpretation, the challenge mentioned by high school students regarding their struggle to prove their maturity and choose a career path was categorized under ``Lack of incentive.'' For the high school group, this category captures the challenges related to the pressure of making career decisions, whereas, for the other groups, it encompasses the lack of encouragement to enter or remain in STEAM fields, often due to insufficient support from colleagues, family, and friends, and broader social discouragement.

\begin{table*}[tphb]
    \centering
    \caption{Comparison of challenges collected from literature and during the workshops}
    \label{tab:comp}
    \begin{tabular}{lp{8cm}|ccc|c|}
        \hline
        \multicolumn{1}{c}{\textbf{Challenge Category}} & \multicolumn{1}{c|}{\textbf{References}} & \textbf{HS} & \textbf{BSc} & \textbf{Pract} & \cellcolor{lightgray} \textbf{Total} \\
        \multicolumn{1}{c}{\textit{(from literature review)}}    & &              &              &                & \cellcolor{lightgray}                \\	\hline
        Inferiorization of women	& \cite{kohl_challenges_2021, trinkenreich_empirical_2022, canedo_breaking_2021, santos_fatores_2023, DBLP:conf/clei/SilvaSM22, sultana_code_2023, wang_implicit_2019, trinkenreich_empirical_2022, longe_literature_2019, araujo_mulheres_2022, DBLP:conf/clei/SilvaSM22} & \cellcolor{yellow}	& \cellcolor{blue}	& \cellcolor{magenta}	  & \cellcolor{lightgray} \textbf{3}	\\	
        Gender bias	 & \cite{longe_literature_2019, DBLP:conf/clei/SilvaSM22, gallindo_critical_2021, dias_canedo_barriers_2019, DBLP:journals/tosem/TrinkenreichWSG22, rodriguez-perez_perceived_2021, trinkenreich_empirical_2022, canedo_breaking_2021, longe_literature_2019, araujo_mulheres_2022, santos_fatores_2023, outao_how_2022, DBLP:conf/clei/SilvaSM22, gallindo_critical_2021, sultana_code_2023}             & \cellcolor{yellow}	& \cellcolor{blue}	& \cellcolor{magenta}	  & \cellcolor{lightgray} \textbf{3}	\\	
        Lack of representation	& \cite{ribeiro_uma_2019, dias_canedo_barriers_2019, wang_implicit_2019, DBLP:journals/tosem/TrinkenreichWSG22, trinkenreich_empirical_2022, canedo_breaking_2021, zhao_workplace_2023, santos_fatores_2023, champa_insights_2023, outao_how_2022, DBLP:conf/clei/SilvaSM22}    &	                    & \cellcolor{blue}  &		                  & \cellcolor{lightgray} \textbf{1}	\\	
        Toxic environment	& \cite{dias_canedo_barriers_2019, DBLP:journals/tosem/TrinkenreichWSG22, trinkenreich_empirical_2022, canedo_breaking_2021, outao_how_2022, DBLP:conf/clei/SilvaSM22, zolduoarrati_value_2021, sultana_code_2023, frluckaj_gender_2022, rodriguez-perez_perceived_2021, zhao_workplace_2023, santos_fatores_2023, tretiakov_social_2023, sultana_code_2023}        & \cellcolor{yellow}	& \cellcolor{blue}	&		                  & \cellcolor{lightgray} \textbf{2}	\\	
        Stereotypes 	& \cite{santos_fatores_2023, kohl_challenges_2021, murphy_examining_2019}            & \cellcolor{yellow}	& \cellcolor{blue}	&		                  & \cellcolor{lightgray} \textbf{2}	\\	
        Overload	        & \cite{dias_canedo_barriers_2019, wang_implicit_2019, DBLP:journals/tosem/TrinkenreichWSG22, murphy_examining_2019, happe_frustrations_2022, longe_literature_2019, araujo_mulheres_2022, santos_fatores_2023, outao_how_2022, tretiakov_social_2023, sultana_code_2023, soares_investigating_2023, tretiakov_social_2023}       &	                    & \cellcolor{blue}  & \cellcolor{magenta}	  & \cellcolor{lightgray} \textbf{2}	\\	
        Feeling of inferiority	 & \cite{happe_frustrations_2022, longe_literature_2019, santos_fatores_2023, frluckaj_gender_2022, DBLP:journals/tosem/TrinkenreichWSG22, trinkenreich_empirical_2022, happe_frustrations_2022, canedo_breaking_2021, outao_how_2022, zolduoarrati_value_2021, frluckaj_gender_2022}   &	                    & \cellcolor{blue}  & \cellcolor{magenta}	  & \cellcolor{lightgray} \textbf{2}	\\	
        Lack of inclusion	& \cite{sultana_code_2023, frluckaj_gender_2022, DBLP:journals/tosem/TrinkenreichWSG22, canedo_breaking_2021, poncell_diversity_2022, canedo_breaking_2021, moro_lack_2022, poncell_diversity_2022, outao_how_2022, tretiakov_social_2023}      & \cellcolor{yellow}    & \cellcolor{blue}	&		                  & \cellcolor{lightgray} \textbf{2}	\\	
        Lack of recognition	  & \cite{kohl_challenges_2021, trinkenreich_empirical_2022, happe_frustrations_2022, canedo_breaking_2021, longe_literature_2019, santos_fatores_2023, sultana_code_2023}      &	                    & \cellcolor{blue}  & \cellcolor{magenta}	  & \cellcolor{lightgray} \textbf{2}	\\	
        Lack of incentive	& \cite{trinkenreich_empirical_2022, happe_frustrations_2022, canedo_breaking_2021, longe_literature_2019, araujo_mulheres_2022, santos_fatores_2023, tretiakov_social_2023, gallindo_critical_2021, canedo_breaking_2021, santos_fatores_2023, DBLP:conf/clei/SilvaSM22}       & \cellcolor{yellow}    &		            &		                  & \cellcolor{lightgray} \textbf{1}	\\	
        Harassment	& \cite{santos_fatores_2023, soares_investigating_2023, tretiakov_social_2023, canedo_breaking_2021, longe_literature_2019, frluckaj_gender_2022}	           & \cellcolor{yellow}    & \cellcolor{blue}	& \cellcolor{magenta}	  & \cellcolor{lightgray} \textbf{3}	\\	
        Career 	    & \cite{dias_canedo_barriers_2019, happe_frustrations_2022, canedo_breaking_2021, longe_literature_2019, trinkenreich_empirical_2022, canedo_breaking_2021, araujo_mulheres_2022, santos_fatores_2023, DBLP:conf/clei/SilvaSM22, sultana_code_2023}                &	                    & \cellcolor{blue}	& \cellcolor{magenta}	  & \cellcolor{lightgray} \textbf{2}	\\	
        Maternity	 & \cite{trinkenreich_empirical_2022, soares_investigating_2023, tretiakov_social_2023, frluckaj_gender_2022}               &	                    &	       	        & \cellcolor{magenta}	  & \cellcolor{lightgray} \textbf{1}	\\ \hline	
        \multicolumn{2}{r|}{\cellcolor{lightgray}\textbf{Total:}} & \cellcolor{lightgray} \textbf{7} & \cellcolor{lightgray}\textbf{11} & \cellcolor{lightgray} \textbf{8} & \multicolumn{1}{c}{} \\ \cline{1-5}
        \multicolumn{6}{p{10cm}}{\begin{footnotesize} \textit{\textbf{Key:} HS = High School Students, BSc = Bachelor Students, and Pract = Practitioners}\end{footnotesize}}
    \end{tabular}
\end{table*}

All three groups reported challenges related to (1) inferiorization of women, (2) gender bias, and (3) harassment. Women, regardless of their life stage, frequently experience being diminished because of their gender. This issue becomes more pronounced as they enter predominantly men environments, such as software development teams or undergraduate software engineering courses. Additionally, all groups reported experiencing gender bias, reflecting societal stereotypes about gender roles. Lastly, cases of harassment—both sexual and moral—were reported across all workshops, including serious incidents that required police intervention.

``Lack of Representation'' was a challenge mentioned exclusively by the bachelor’s degree students. It is understandable that high school students did not mention this issue, as gender balance is more or less maintained at that stage. The absence of this challenge in the practitioners' discussions does not necessarily indicate that it is unimportant; instead, it suggests that other challenges may have been more pressing for this particular group.

Another challenge that surfaced only among the practitioners was related to maternity, which is logical, given that most women typically consider starting a family during the stage represented by the practitioners. Interestingly, the bachelor’s student group identified the greatest number of challenges, which could be attributed to the larger number of participants in this group and the comfort level they exhibited in sharing their experiences.

This research also investigates solutions to the challenges faced by women in Software Engineering and compares the solutions reported in the literature with those collected during our workshops, as illustrated in Table \ref{tab:compSol}. 
\begin{table*}[bpht]
    \centering
    \caption{Comparison of solutions collected from literature and during the workshops}
    \label{tab:compSol}
    \begin{tabular}{lp{8cm}|ccc|c|}
        \hline
        \multicolumn{1}{c}{\textbf{Solution Category}} & \multicolumn{1}{c|}{\textbf{References}} & \textbf{HS} & \textbf{BSc} & \textbf{Pract}      & \cellcolor{lightgray} \textbf{Total} \\
        \multicolumn{1}{c}{\textit{(from literature review)}}     & &                      &                         &                     & \cellcolor{lightgray}                \\	\hline
        Awareness initiatives &	\cite{wang_implicit_2019,rodriguez-perez_perceived_2021,happe_frustrations_2022, zhao_workplace_2023, outao_how_2022, poncell_diversity_2022,DBLP:journals/tosem/TrinkenreichWSG22, moro_lack_2022,trinkenreich_empirical_2022}	        &		             &	\cellcolor{blue}    & \cellcolor{magenta}	  &	\cellcolor{lightgray} \textbf{2}	\\
        Affirmative action &	\cite{kohl_challenges_2021, dias_canedo_barriers_2019, trinkenreich_empirical_2022, soares_investigating_2023, canedo_breaking_2021, tretiakov_social_2023, sultana_code_2023, longe_literature_2019, poncell_diversity_2022,araujo_mulheres_2022, santos_fatores_2023}	            &		             &		                & \cellcolor{magenta}	  & \cellcolor{lightgray} \textbf{1}	\\
        Policies to encourage the presence of &	\cite{canedo_breaking_2021, santos_fatores_2023, gallindo_critical_2021, araujo_mulheres_2022, longe_literature_2019, moro_lack_2022, dias_canedo_barriers_2019, DBLP:journals/tosem/TrinkenreichWSG22, trinkenreich_empirical_2022, murphy_examining_2019,rodriguez-perez_perceived_2021} &	&	& &	\cellcolor{lightgray} \textbf{0} \\
        women in STEM fields &	 &	&	& &	\cellcolor{lightgray}\\
        Strategies to combat gender bias & \cite{DBLP:journals/tosem/TrinkenreichWSG22, rodriguez-perez_perceived_2021, soares_investigating_2023, poncell_diversity_2022, dias_canedo_barriers_2019,araujo_mulheres_2022, rodriguez-perez_perceived_2021, tretiakov_social_2023,longe_literature_2019,DBLP:conf/clei/SilvaSM22, sultana_code_2023, outao_how_2022}	                        &		             &	\cellcolor{blue}	&		                  &	\cellcolor{lightgray} \textbf{1}	\\
        Combating harassment &	\cite{dias_canedo_barriers_2019, rodriguez-perez_perceived_2021}	        &		             &		                & \cellcolor{magenta}     & \cellcolor{lightgray} \textbf{1}	\\
        Promoting representativeness &	\cite{DBLP:journals/tosem/TrinkenreichWSG22, longe_literature_2019, sultana_code_2023, tretiakov_social_2023,wang_implicit_2019, rodriguez-perez_perceived_2021, trinkenreich_empirical_2022, soares_investigating_2023, araujo_mulheres_2022,canedo_breaking_2021, zolduoarrati_value_2021}   &		             &	\cellcolor{blue}	&		                  &	\cellcolor{lightgray} \textbf{1}	\\
        Supporting women &	\cite{trinkenreich_empirical_2022, tretiakov_social_2023, soares_investigating_2023, santos_fatores_2023,DBLP:journals/tosem/TrinkenreichWSG22,zhao_workplace_2023,longe_literature_2019,frluckaj_gender_2022}	            & \cellcolor{yellow} &		                & \cellcolor{magenta}     & \cellcolor{lightgray} \textbf{2}	\\
        Inclusive conduct &	\cite{trinkenreich_empirical_2022, rodriguez-perez_perceived_2021, frluckaj_gender_2022, outao_how_2022, tretiakov_social_2023, zhao_workplace_2023, DBLP:journals/tosem/TrinkenreichWSG22, longe_literature_2019, soares_investigating_2023, sultana_code_2023, dias_canedo_barriers_2019}	            &		             &		                &		                  & \cellcolor{lightgray} \textbf{0}	\\ \hline
          \multicolumn{2}{r|}{\cellcolor{lightgray}\textbf{Total:}} &	\cellcolor{lightgray} \textbf{1}	&	\cellcolor{lightgray} \textbf{3}      &	\cellcolor{lightgray} \textbf{4}	&	\multicolumn{1}{c}{}\\ \cline{1-5}
        \multicolumn{5}{p{10cm}}{\begin{footnotesize} \textit{\textbf{Key:} HS = High School Students, BSc = Bachelor Students, and Pract = Practitioners}\end{footnotesize}}
    \end{tabular}
\end{table*}
The first column of Table \ref{tab:compSol} lists the categories of solutions followed by the studies where we collected them. The subsequent columns compare these with the findings from the workshops. 

``Awareness Initiatives'' and ``Supporting Women'' were the workshops' most frequently mentioned solution categories. Both undergraduate students and professionals emphasized the need for discussion and training on gender diversity, with students noting that men's participation in these discussions would be particularly beneficial. The category ``Supporting Women'' included suggestions from high school students that support from adults would help them achieve their goals, even though the support was not exclusively for women. For professionals, support was also suggested in terms of assistance with childcare expenses.

Despite being prominent in the literature, the category ``Inclusive Conduct'' was not mentioned by any participants during the workshops. This category addresses challenges related to fostering respectful attitudes and active voices for women in their teams and organizations. Another overlooked category was ``Policies to Encourage the Presence of Women in STEM,'' which includes solutions aimed at increasing women's entry into and retention in STEM fields. These initiatives are crucial not only for addressing immediate challenges but also for fostering a cultural shift towards greater gender diversity in STEM. The effectiveness of these policies lies in their ability to inspire young women to pursue careers in software engineering and provide them with the necessary support to overcome obstacles.

In summary, our research highlights that many challenges women face in entering and advancing within the software development field are deeply rooted in societal biases, as well as structural and cultural barriers that persist at various stages of their professional journeys. While high school students, undergraduate students, and practitioners experience similar forms of gender-based challenges, the specific contexts and pressures they face vary across life stages, as reflected in our findings. Importantly, our study not only underscores the necessity of addressing immediate issues such as harassment and gender bias but also calls attention to the broader, systemic changes required to foster long-term gender diversity and inclusion in the field.

\section{Conclusions} 
\label{sec:conl}
This study aimed to explore women's challenges across three distinct life stages: high school, University, and professional practice. We began by conducting a literature review to identify existing challenges and then presented these findings for discussion among three groups of women. Through this research, we systematically organized the challenges women encounter and compared them across different life stages. This comparison revealed that certain challenges, such as inferiority and harassment, are recurrent regardless of the profession or age. Other challenges, like maternity, are specific to certain stages. Additionally, some challenges emerge only after women have chosen a career in software development.

From a practical point of view, our findings can be used to improve women's inclusion and experience in the IT field. Understanding and addressing these challenges can help create a more supportive environment for women, encouraging greater participation and retention. From a research perspective, the challenge categories provide a framework for further studies, and our research design can serve as a model for similar investigations.

However, the development of this project was not without difficulties. The primary challenge was the low number of participants at the workshops, which limited the diversity and size of our sample. Additionally, we found that discussing these sensitive issues required more time and effort than anticipated, as participants needed time to feel comfortable and open up about their experiences.

The primary threats to the validity of this research stem from the variability in the environment, participant characteristics, and data collection methods across the three workshops. The workshop had different settings (school, university campus, online), which could have influenced participants' engagement and responses, and because the sessions were not recorded, it may have led to incomplete or subjective interpretation. Also, the workshops were conducted in a specific sequence, which could introduce temporal bias. Finally, our workshop is limited by its sample size and sampling strategies, restricting the ability of generalizations. 

For future work, conducting additional workshops with a larger and more diverse sample would be valuable. Furthermore, research should explore other dimensions of social diversity, including race, culture, disabilities, and sexuality, as these are critical factors influencing the challenges individuals encounter in various aspects of life. This would help clarify which challenges are most prevalent at each life stage and increase the generalizability of our findings. 

\section*{Acknowledgment}

We are grateful to everyone who participated in this research, including high school students, university students, and practitioners, for sharing their stories and experiences. We also extend our special thanks to the teachers and the head of the high school we visited for graciously allowing us to conduct our research within their institution. The University of Brasilia supported this research through the DEX/DPI/SDH No. 05/2023 funding.


\end{document}